\documentclass[twocolumn,showpacs,prb,floats,epsfig,superscriptaddress]{revtex4}
\usepackage{epsfig}
\usepackage{graphicx}
\usepackage{amsfonts}
\usepackage{amsmath}
\usepackage{amssymb}
\usepackage{amsbsy}
\usepackage[dvips]{color}
\usepackage{color}

\newcommand{\beq}{\begin{equation}}
\newcommand{\eeq}{\end{equation}}
\newcommand{\bea}{\begin{eqnarray}}
\newcommand{\eea}{\end{eqnarray}}

\newcommand{\bra}[1]{\langle #1|}
\newcommand{\ket}[1]{|#1\rangle}

\begin{document}

\title{Dynamic freezing and defect suppression in the tilted one-dimensional Bose-Hubbard model}

\author{U. Divakaran}
\affiliation{Department of Physics, Indian Institute of Technology,
Kanpur, India.}

\author{K. Sengupta}
\affiliation{Theoretical Physics Department, Indian Association for
the Cultivation of Science, Jadavpur, Kolkata-700032, India.}

\begin{abstract}
We study the dynamics of tilted one-dimensional Bose-Hubbard model
for two distinct protocols using numerical diagonalization for
finite sized system ($N\le 18$). The first protocol involves
periodic variation of the effective electric field $E$ seen by the
bosons which takes the system twice (per drive cycle) through the
intermediate quantum critical point. We show that such a drive leads
to non-monotonic variations of the excitation density $D$ and the
wavefunction overlap $F$ at the end of a drive cycle as a function
of the drive frequency $\omega_1$, relate this effect to a
generalized version of St\"uckelberg interference phenomenon, and
identify special frequencies for which $D$ and $1-F$ approach zero
leading to near-perfect dynamic freezing phenomenon. The second
protocol involves a ramp of both the electric field $E$ (with a rate
$\omega_1$) and the boson hopping parameter $J$ (with a rate
$\omega_2$) to the quantum critical point. We find that both $D$ and
the residual energy $Q$ decrease with increasing $\omega_2$; our
results thus demonstrate a method of achieving near-adiabatic
protocol in an experimentally realizable quantum critical system. We
suggest experiments to test our theory.

\end{abstract}

\maketitle

\section{Introduction}

Ultracold atom systems, in the presence of optical lattices, have
proved to be successful emulators of several model Hamiltonians such
as the Ising and the Bose-Hubbard models \cite{bloch1}. These
systems offer unprecedented tunability of the parameters of the
Hamiltonians they emulate. Consequently, they serve as perfect test
bed for studying the low-temperature properties and possible quantum
phase transitions of the emulated models \cite{bloch2}. Furthermore,
ultracold atom systems provide a near-perfect isolation of its
constituents from the environment; this feature, along with
real-time tunability of the laser intensity used to create the
optical lattice, makes them ideal systems for studying
non-equilibrium dynamics of a closed quantum system near its
critical point \cite{rev1}. Several such studies, both experimental
and theoretical, have already been undertaken for a system of
ultracold bosons in an optical lattice emulating the Bose-Hubbard
model near a Mott-insulator (MI)- superfluid (SF) quantum critical
point \cite{dupuis1,trefzger1,mft1,bloch2,greiner1}.

More recently, following the theoretical prediction of Ref.\
\onlinecite{sachdev1}, there has been experimental realizations of a
translational symmetry broken density wave ground state of
one-dimensional (1D) ultracold bosons in the presence of an
effective electric field \cite{bakr1}. The atoms in the trap are
neutral; thus such an electric field can be generated either by
shifting the center of the confining trap \cite{bloch2} or by
applying a spatially varying Zeeman magnetic field \cite{bakr1}. It
is well-known that the low-energy properties of such a system of
bosons can be described by an effective dipole model with a
Hamiltonian \cite{sachdev1}
\begin{eqnarray}
H_d=-J \sqrt{n_0(n_0+1)} \sum_{\ell}
(d_{\ell}^{\dagger}+d_{\ell})+(U-E) \sum_{\ell} n_{\ell},
\label{diham1}
\end{eqnarray}
where $d_{\ell}=b_i^{\dagger} b_j$ is the dipole annihilation
operator living on the link $\ell$ between sites $i$ and $j$. Such a
dipole constitutes a bound state of a boson and a hole on adjacent
lattice sites $i$ and $j$. Here $b_i$ is the boson annihilation
operator on site $i$, $U$ is the on-site repulsion of the bosons,
$n_0$ is the number of bosons in the MI phase at each lattice site,
$J$ is their hopping amplitude, $n_{\ell}=d_{\ell}^{\dagger}
d_{\ell}$ is the dipole number operators residing on the link
$\ell$, and $E$ is the effective electric field. The dipole
operators satisfy additional constraints of having a maximum of
single dipole per link ($n_{\ell} \le 1$) and having at most one
dipole on two adjacent links ($n_{\ell} n_{\ell \pm 1}=0$). These
dipole states are resonantly coupled to the parent Mott state when
$U \sim E$. It has been shown that such a system undergoes a phase
transition from a dipole vacuum state for $U\gg E$ to the one with
maximum number of dipoles ($=N/2$, where $N$ denotes the chain
length in units of lattice spacing) when $E \gg U$. This phase
transition belongs to the Ising universality class as reported in
Ref. \onlinecite{sachdev1} and the critical point $E_c$ is given by
$E_c=U+1.31J\sqrt{n_0(n_0+1)}$. It turns out that such a system of
bosons also has a description in terms of Ising spins and
constitutes the realization of an effective Ising model with both
longitudinal ($\sim (U-E)$) and transverse ($\sim J$) magnetic
fields \cite{sachdev1,bakr1}. In the spin language, the dipole
vacuum and the maximal dipoles states are termed as paramagnetic
(PM) and Ising antiferromagnetic (AFM) respectively \cite{bakr1}.
Such studies have recently been extended to include the effect of
higher dimensions \cite{sachdev2}, disorder \cite{sachdev3}, and for
weakly coupled bosons \cite{manmana1}.

In the last few years, several studies have been performed to
understand different aspects of non-equilibrium dynamics of closed
quantum systems. Most of the initial studies in this direction
focussed either on sudden quenches \cite{sengupta1,cardy1,
calabrese1} or on linear or non-linear ramp protocols taking the
systems through quantum critical points \cite{zureck1,anatoli1,
sen1,sengupta2,dutta1,mukherjee07,anatoli2}. The former class of
studies were mostly concerned with the evolution and long-time
behavior of a closed quantum system following a quench while the
latter group have demonstrated the possibility of realization of
Kibble-Zurek scaling \cite{kibble1, zureck2} and its variants in the
context of isolated quantum systems. More recently, several studies
have focussed on periodic dynamics of closed quantum systems; such a
dynamics inevitably involve multiple passage of a quantum system
through the intermediate quantum critical point which leads to novel
interference phenomenon \cite{mukherjee09,nori1}. Moreover, such
protocols lead to the realization of dynamic freezing of the state
of the system during a periodic drive \cite{das1, mondal1} and may
lead to novel steady states \cite{das2}. However, most of these
studies have not been applied to experimentally realizable
non-integrable systems. Further, it has also been recently shown
that for a generic quantum Hamiltonian, a two-parameter drive
protocol, which constitutes a time dependent ramp of two of the
Hamiltonian parameters, may lead to suppression of defects during
the passage of a system through a quantum critical point
\cite{sau1}. However, such a protocol has never been applied to a
specific experimentally relevant and/or non-integrable model.

In this work, we study the non-equilibrium dynamics of the dipole
Hamiltonian $H_d$ for two separate (periodic and two-rate) protocols
using numerical diagonalization for finite sized system ($N\le 18$).
The former protocol involves periodic variation of the effective
electric field $E$ seen by the bosons with a drive frequency
$\omega_1$ while the latter involves linear ramp of $E$ the and the
boson hopping $J$ with rates $\omega_1$ and $\omega_2$ respectively.
We note at the outset that the quench and the ramp dynamics of this
model has been studied in Refs.\ \onlinecite{sengupta1} and
\onlinecite{kolo1}. The former study has predicted that the
long-time average of the dipole order parameter will have a maximal
value when the final value of the electric field after the ramp,
starting from the dipole vacuum phase, matches with $U$. The latter
studies showed that these system could prove as experimental test
bed for realization of Kibble-Zurek law for finite size systems.
However, the behavior of this model when driven by periodic or
two-rate protocols has never been studied. In this work, we aim to
fill up this gap.

The main results of our work are as follows. First, we show that a
periodic variation of $E$ with a rate $\omega_1$, which takes the
system twice through the intermediate critical point for each drive
period starting from the dipole vacuum phase, leads to non-monotonic
variation of the excitation (defect) density $1-F$ (where $F$ is the
overlap between the state of the system and the instantaneous ground
state at the end of the drive), and the dipole excitation density
$D$, measured after an integer number of drive cycles, as a function
of $\omega_1$. Second, we identify specific frequencies for which
$(1-F), D \sim 0$ after a complete drive cycle leading to
near-perfect dynamic freezing phenomenon \cite{das1,mondal1}.  We
also show that both these phenomena occurs due to the quantum
interference effect originating from multiple passage of the system
through the quantum critical point and constitute an example of
generalized form of St\"uckelberg interference phenomenon
\cite{nori1}. Third, for the two-rate protocol, we demonstrate that
$D$ and the residual energy $Q$ exhibit power law dependence on both
$\omega_1$ and $\omega_2$ over a range of drive frequency; they
increase (decrease) with increasing $\omega_1$ ($\omega_2$). We
identify the corresponding exponents and compare them with the
prediction of Kibble-Zurek theory for finite sized system
\cite{kolo1}. Finally, we chart out the range of drive frequencies
where one expects to experimentally observe such behavior for
experimentally relevant finite sized system. We point out that owing
to the decrease of $D$ and $Q$ with $\omega_2$, such a two-rate
protocol may be near-adiabatic and thus might be useful for quantum
state preparation in an experimentally realizable system upon its
passage through a quantum critical point. We note that both the
non-monotonic behavior of $D$ and $F$ for the periodic protocol and
the suppression of $D$ and $Q$ with increasing $\omega_2$ provide
examples of phenomena that have no analog in standard quench and
ramp protocols; thus we expect our results to provide additional
relevant information for possible future experiments in this system.

The plan of the rest of the paper is as follows. In Sec.\
\ref{method}, we chart out the details of the protocols studied and
the method of our analysis. This is followed by Sec.\ \ref{result},
where we present the main results for periodic and two rate
dynamics. Finally we conclude with a discussion of our main results
and their experimental relevance in Sec.\ \ref{conc}.

\section{Analysis of the tilted Bose Hubbard model}
\label{method}

The boson Hubbard model, in the presence of an effective electric
field $E$, which essentially tilts the lattice along one direction,
can be written as \bea H &=&-J\sum_{\langle ij \rangle}
(b_i^{\dagger}b_j + b_j^{\dagger}b_i)+\frac{U}{2}\sum_j n_j(n_j-1)
\nonumber\\ && -E\sum_j j n_j. \label{bhoriginal} \eea Here
$n_j=b_j^{\dagger}b_j$ denotes the number of bosons at site $j$, and
$U$ is the on-site repulsive interaction potential. As shown in Ref.
\onlinecite{sachdev1}, starting from the parent Mott state with
$n_0$ bosons per site, the low energy behavior of the tilted Bose
Hubbard model, for $U,E \gg |U-E|,J$, can be described by the
effective dipole Hamiltonian $H_d$ given by Eq.\ \ref{diham1}. In
what follows we shall analyze the dynamics of the dipoles in the
presence of time-dependent hopping strength $J(t)$ and electric
field $E(t)$ such that $U,E(t) \gg |U-E(t)|,J(t)$ at all times so
that $H_d$ can be reliably used to describe the dynamics of the
dipoles. We shall use numerical exact diagonalization technique for
obtaining eigenvalues and eigenfunctions of $H_d$; this limits the
size of the system to $N\le 18$. We note such system sizes are
similar to what has been experimentally achieved in Ref.\
\onlinecite{bakr1}; thus our results are expected to be of direct
relevance to possible future experiments on these systems.

We first consider the periodic protocol for which
\begin{eqnarray}
E(t)= U - E_0 \cos(\omega_1 t), \label{ppcol1}
\end{eqnarray}
where $E_0$ is chosen such that the system starts in a dipole vacuum
state at $t=0$. Note that with this choice of the protocol, the
instantaneous energy of dipole formation, $U-E(t)$, vanishes twice
at $ \omega_1 t= \pi/2, 3 \pi/2$ during each drive cycle. Also, the
system crosses critical point when $E(t)=E_c$; this also occurs
twice at $t=t_{1,2}$ for each drive cycle, where
\begin{eqnarray}
\omega_1 t_1 &=& \cos^{-1}(-\mu_0J/E_0), \quad t_2 = 2\pi/\omega_1
-t_1, \label{crosstimes}
\end{eqnarray}
where $\mu_0=1.3 \sqrt{n_0(n_0+1)}$. The Schr\"odinger equation for
the many-body wavefunction $|\psi\rangle$ in the presence of a
periodic $E(t)$ is given by
\begin{eqnarray}
i \hbar \partial_t |\psi(t)\rangle &=& H_d [E(t)] |\psi(t)\rangle
\label{sch1}
\end{eqnarray}
To solve Eq.\ \ref{sch1}, we expand $|\psi(t)\rangle = \sum_n
c_{\alpha}(t) |\alpha \rangle$, where $|\alpha\rangle$ denotes the
instantaneous eigenstates of $H_d$ for $E= E(t=0)$. These
eigenstates satisfy $H_d(t=0)|\alpha \rangle = \epsilon_{\alpha}
|\alpha \rangle$, where $\epsilon_{\alpha}$ denotes the
instantaneous eigenenergies at $t=0$. Here the coefficients
$c_{\alpha}(t)$ represent the overlap of the state $|\psi(t)\rangle$
with $|\alpha\rangle$. Eq.\ \ref{sch1} can now be reexpressed as
coupled set of differential equations governing the time evolution
of $c_{\alpha}(t)$. These equations are given by
\begin{eqnarray}
(i \hbar \partial_t - \epsilon_{\alpha} )c_{\alpha} &=& E_0
[1-\cos(\omega_1 t)] \sum_\beta \Lambda_{\alpha \beta}^{(1)}
c_{\beta}(t), \nonumber\\
\Lambda_{\alpha \beta}^{(1)} &=& \langle \alpha |\sum_{\ell}
n_{\ell} |\beta\rangle \label{ceq}
\end{eqnarray}
with the initial condition $c_{\alpha}(0)=c_{0 \alpha}$. The
coefficient $c_{0 \alpha}$, and the eigenenergies
$\epsilon_{\alpha}$ are obtained by exact diagonalization of
$H_d(t=0)$. A numerical solution of these equations yields the state
of the system $|\psi(t)\rangle$ at any time $t$ during the drive.

Having obtained $|\psi(t)\rangle$, one can use it to compute
expectation values of several relevant quantities. In the present
work, we shall mainly concentrate on the wavefunction overlap $F$,
the dipole excitation density $D$, and the residual energy $Q$ of
the system. In terms of the overlap coefficients $c_{\alpha}(t)$,
one can obtain these quantities as
\begin{eqnarray}
D(t)&=&  |n_d(t)-\langle \psi_G(t)|\sum_{\ell} n_{\ell}| \psi_G(t)
\rangle/N|  \nonumber\\
&=& |n_d(t)-n_d^G(t)| \\
n_d(t) &=& \frac{1}{N}\langle \psi(t)|\sum_{\ell} n_{\ell}
|\psi(t)\rangle = \frac{1}{N}
\sum_{\alpha \beta} c_{\alpha}^{\ast}(t) c_{\beta}(t) \Lambda_{\alpha \beta}^{(1)} \nonumber\\
F(t) &=& |\langle\psi(t)|\psi_G(t)\rangle|^2,\nonumber\\
Q(t)&=&\bra{\psi(t)}H(t)\ket{\psi(t)}-\bra{\psi_G(t)}H(t)\ket{\psi_G(t)}\nonumber\\
&=& \sum_{\alpha \beta} c_{\alpha}^{\ast}(t) c_{\beta}(t) \langle
\alpha| H(t) |\beta\rangle- \bra{\psi_G(t)}H(t)\ket{\psi_G(t)},
\nonumber\ \label{def1}
\end{eqnarray}
where $|\psi_G(t)\rangle$ is the instantaneous ground state of the
system at time $t$, $n_d$ is the dipole density at time $t$, $n_d^G$
is the dipole density corresponding to the instantaneous ground
state at $t$, and we have set the lattice spacing to unity. We note
that the expressions of $n_d$, $D$, $F$ and $Q$ obtained in Eq.\
\ref{def1} assume a particularly simple form when evaluated at the
end of an integer ($p$) number of drive cycles, {\it i.e.}, at
$t=t_f= 2\pi p/\omega_1$. This simplicity arises from the fact that
$H(t_f)=H(0)$ in these cases and leads to
\begin{eqnarray}
n_d(t_f) &=& \frac{1}{N} \sum_{\alpha \beta} c_{\alpha}^{\ast}(t_f)
c_{\beta}(t_f) \Lambda_{\alpha \beta}^{(1)}, \quad F(t_f) =
|c_0(t_f)|^2 \nonumber\\
D(t_f)&=& |n_d(t_f)- \Lambda_{00}^{(1)}/N|,
\nonumber\\
Q(t) &=& \sum_{\alpha \ne 0} (\epsilon_{\alpha} -\epsilon_0)
|c_{\alpha}(t_f)|^2, \label{percoeff}
\end{eqnarray}
where we have denoted the initial ground state of the system by
$|\alpha=0\rangle \equiv |0\rangle$.

Finally, we consider the two rate protocol where one varies both the
boson hopping amplitude $J$ and the effective electric field $E$
according to the protocol
\begin{eqnarray}
E(t)&=& E_i + \Delta E \omega_1 t, \quad  J(t)= \epsilon + \Delta J
\omega_2 t \label{prototwo}
\end{eqnarray}
where the initial value of the electric field $E_i$ is chosen so
that the system is in the paramagnetic phase, and $\epsilon/\Delta
J$ is chosen to be a small number. The ramp starts at $t=0$ and
continues till $t=t_c$ when the system reaches the critical point:
$E(t_c)=U+ \mu_0 J(t_c)$. This yields
\begin{eqnarray}
t_c &=& (U-E_i -\mu_0\epsilon)/\omega_0, \nonumber\\
\omega_0 &=& (\omega_1 \Delta E - \mu_0 \omega_2 \Delta J).
\label{tcdef}
\end{eqnarray}
We note that for $\omega_2 > \omega_2^c = \omega_1 \Delta E/(\mu_0
\Delta J)$, $\omega_0 <0$ leading to $t_c <0$ which indicates that
for fast enough $\omega_2$ the system is not going to reach the
critical point. In this work, we shall restrict ourselves to
$\omega_2 \le \omega_2^c$.

To obtain the solution of Eq.\ \ref{sch1}, we follow a procedure
similar to the case of the periodic single parameter drive and
expand the wavefunction $|\psi(t)\rangle = \sum_{\alpha} c'_{\alpha}
(t) |\alpha\rangle$, where $|\alpha\rangle$ denotes the eigenstates
of $H(t=0)$ with $E=E_i$ and $J=\epsilon$ satisfying
$H|\alpha\rangle = \epsilon'_{\alpha} |\alpha\rangle$. Eq.\
\ref{sch1} then leads to the coupled equations of motion for
$c'_{\alpha}(t)$
\begin{eqnarray}
(i \hbar \partial_t -\epsilon'_{\alpha}) c'_{\alpha}(t) &=& -t
\sum_{\beta} (\Delta E \omega_1  \Lambda^{(1)}_{\alpha \beta}
+\Delta J \omega_2 \Lambda_{\alpha \beta}^{(2)} ) c'_{\beta} (t)
\nonumber\\
\Lambda^{(2)}_{\alpha \beta} &=& \langle \alpha | \sum_{\ell}
(d_{\ell}+ d_{\ell}^{\dagger})|\beta\rangle \label{cprimeeq}
\end{eqnarray}
with the initial condition $c'_{\alpha}(t=0)=c'_{0 \alpha}$. These
equations can be solved numerically and $c'_{0 \alpha}$ and
$\epsilon'_{\alpha}$ can be obtained by exact diagonalization of
$H_d(t=0)$. This procedure leads to $|\psi(t)\rangle$ and hence, via
Eq.\ \ref{def1}, to $n_d$, $Q$, $D$ and $F$.

\section{Results}
\label{result}

In this section, we discuss the results obtained by numerical
analysis of Eq.\ \ref{ceq} and \ref{cprimeeq}. In Sec.\ \ref{ppcol},
we discuss our results involving compuattion of $D$ and $F$ for the
periodic protocol (Eq.\ \ref{ppcol1}). This is followed by Sec.\
\ref{trcol}, where we numerically compute $D$ and $Q$ for two-rate
protocol (Eq.\ \ref{prototwo}).

\subsection{ Periodic Protocol}
\label{ppcol}

In the presence of the periodic drive (Eq.\ \ref{ppcol1}), the
instantaneous ground state of the tilted Bose Hubbard model changes
from zero dipole (PM) to the maximum dipole (AFM) state and back,
passing twice through the intermediate quantum critical point for
each drive cycle. For this protocol, $\omega_1 \to 0$ corresponds to
the adiabatic limit where system remains close to the instantaneous
ground state at all times.

\begin{figure}[h]
\includegraphics[height=2.6in]{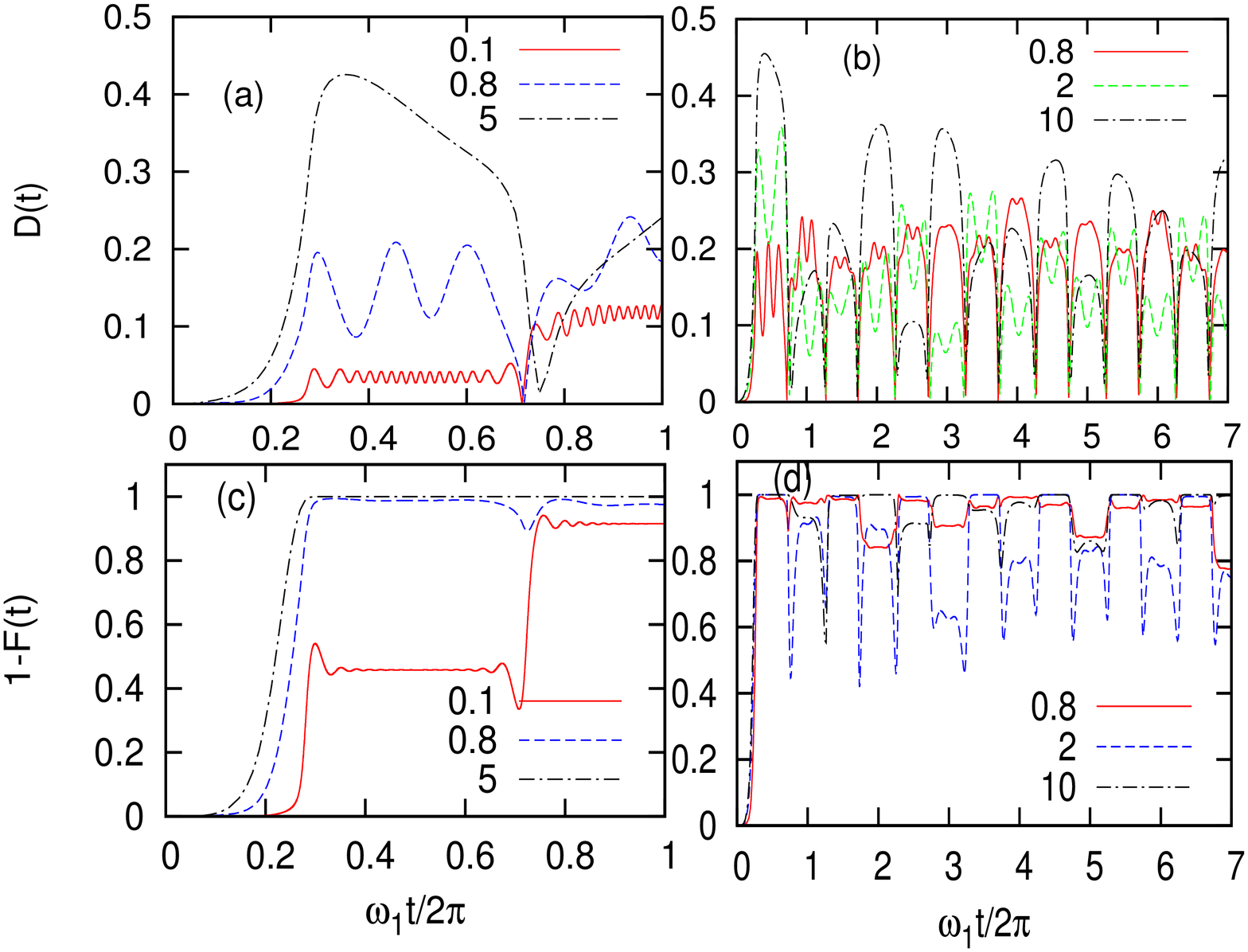}
\caption{(a) and (b): Plot of variation of the dipole excitation
density $D(t)$ as a function of $\omega_1 t$ for different scaled
frequency $\omega_1/(E_c-U)$ (shown in the figure) up to one (a) and
seven (b) drive periods. (c) and (d): Plots of the defect density
$1-F(t)$ as a function of $\omega_1 t$ for same parameters as (a)
and (b) respectively. } \label{fig_D_F}
\end{figure}

\begin{figure}[h]
\includegraphics[width=2.2in,angle=-90]{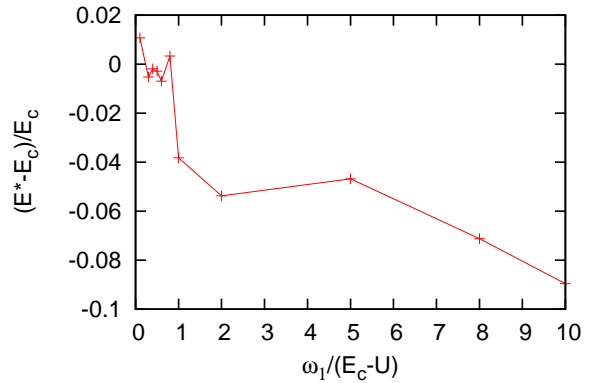}
\caption{Plot of the position of the dip in $D$,
$(E^{\ast}-E_c)/E_c$, as a function of scaled frequency
$\omega_1/(E_c-U)$. The dip is closer to the critical point for
smaller frequencies. } \label{dip}
\end{figure}

We first present our results obtained through numerical simulations
as described in the previous section for $(U-E_0)/J=30$. In Fig.\
\ref{fig_D_F}(a), we plot the time evolution of dipole excitation
density $D$ as a function of time $t$ for one complete cycle and for
several scaled frequencies $\omega_1/(E_c-U)$. As can be seen from
the figure, and as theoretically expected, fewer defects are
generated during the drive for smaller frequencies. For a fixed
frequency, $D = |n_d(t)-n_d^G(t)|$ starts increasing when the
critical region is crossed for the first time (around $t=t_1$). This
happens since for small frequencies, the system enters the impulse
region around the critical point where the state of the system
starts to deviate from the instantaneous ground state (which, for
$E\le E_c$ has $n_d^G(t) \simeq 0$) leading to increase in $D$. As
we continue the evolution within the AFM phase, the system stays in
an excited state with $n_d(t)<n_d^G(t)$ for $t_1<t<t_2$, with higher
$D(t)$ for larger frequencies. After crossing the critical point for
the second time and reaching the PM phase, the reverse condition is
true with $n_d(t)>n_d^G(t) (\sim 0)$. This leads to a dip in $D(t)$
when $n_d(t)=n_d^G(t)$. We find numerically that in the periodic
case, the dip in $D(t)$ occurs at $E=E^{\ast} \simeq E_c$,
especially for small frequencies as shown in Fig.\ \ref{dip}. We
note however, that the fact $D \simeq 0$ does not mean that the
state of the system is identical to the instantaneous ground state
at this point and does not constitute an example of dynamical
freezing. This can be seen from a plot of $1-F$ as a function of
$t$; we find that $1-F$ does not approach zero concomitantly with
$D$. The corresponding behavior of $1-F$ as a function of $t$ is
shown in Fig.\ \ref{fig_D_F}(c).

\begin{figure}[h]
\includegraphics[width=3.5in]{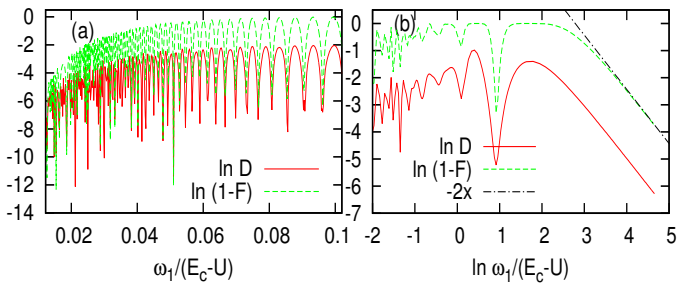}
\caption{Plot of $\ln(D)$ and $\ln(1-F)$ after one complete cycle as
a function of $\omega_1$. (a) corresponds to small frequency region
where for certain frequencies, $F$ is close to unity (or $1-F \sim
0$) which signals near perfect freezing of the wavefunction after
one complete cycle. As the frequency increases, the freezing
phenomena becomes less effective and ultimately disappears for
$\ln[\omega_1/(E_c-U)]> 2$ as shown in (b). } \label{fig_fullcycle}
\end{figure}

Next, we study the behavior of $D$ and $1-F$ after a complete cycle
of drive as a function of $\omega_1$. The results are shown in Fig.\
\ref{fig_fullcycle}. We find that both $D$ and $1-F$, after a
complete drive cycle, display non-monotonic oscillatory behavior as
a function of $\omega_1$. Also, as can be seen in Fig
\ref{fig_fullcycle}(a), there are certain special frequencies at
which $1-F(T)$ and $D(T)$, where $T=2\pi/\omega_1$, concomitantly
approach zero signalling near perfect revival of the wavefunction.
This phenomenon is termed as dynamics induced freezing in Ref.\
\onlinecite{mondal1}. To explore how close one approaches near
perfect freezing in the present system, we plot $\ln D \equiv \ln
D(T)$ and $\ln(1-F) \equiv \ln[1-F(T)]$ as a function of $\omega_1$
in Fig. \ref{fig_fullcycle}. As can be seen in Fig
\ref{fig_fullcycle}(a), $\ln(1-F)$ and $\ln D$ can be as low as
$-10$ for small $\omega_1$. In Fig.\ \ref{fig_fullcycle}(b), one
finds that the freezing is effective for $\omega_1/|U-E_c| \ll 1$;
for larger frequencies, the freezing phenomenon disappears and $1-F$
and $D$ decreases monotonically with $\omega_1$. In particular, in
the large frequency regime, both $D, (1-F) \sim \omega_1^{-2}$.

\begin{figure}[h]
\includegraphics[width=0.7 \columnwidth,angle=-90]{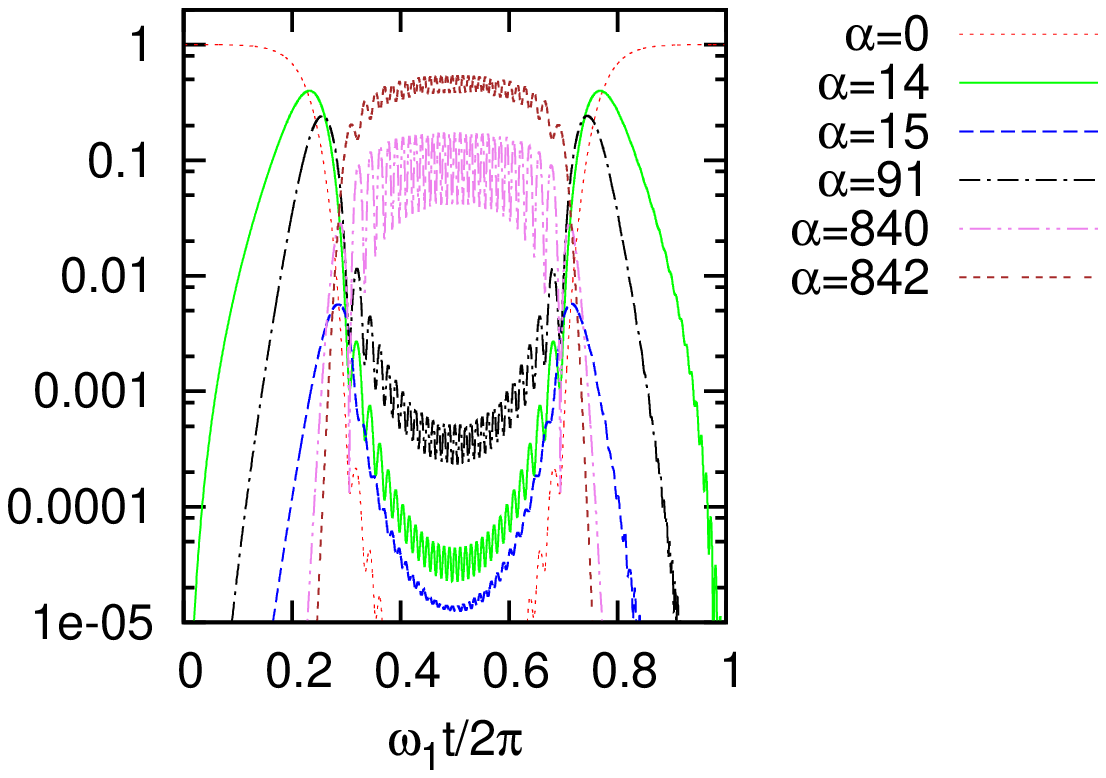}
\includegraphics[width= 0.7 \columnwidth,angle=-90]{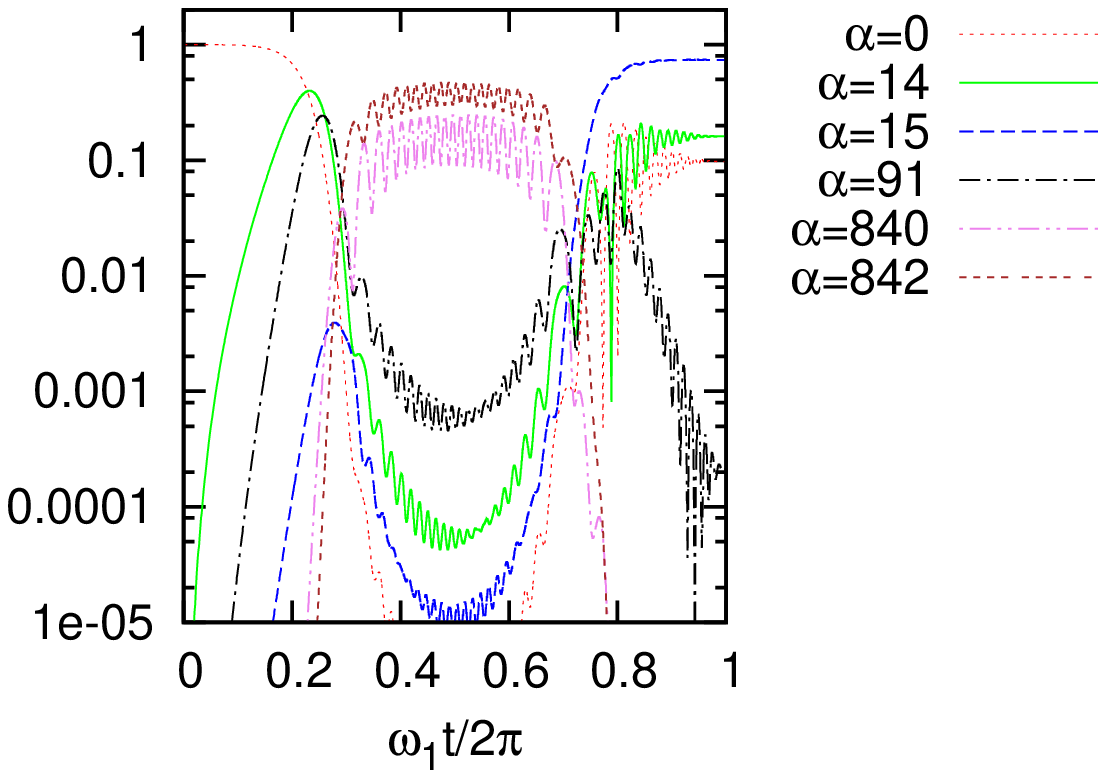}
\caption{Plot of the wavefunction overlap coefficients
$|c_{\alpha}(t)|^2$ for selected set of $\alpha$ satisfying the
condition $|c_{\alpha}(t)|^2\ge 1E-5$ as a function of $\omega_1 t$
during the drive cycle. The top figure corresponds to
$\omega_1/(E_c-U))=0.051$ corresponding to a minima of $D$ while
that at the bottom has $\omega_1/(E_C-U)=0.074$ corresponding to a
maxima of $D$.} \label{fig4}
\end{figure}

\begin{figure}[h]
\includegraphics[width=2.0in,angle=-90]{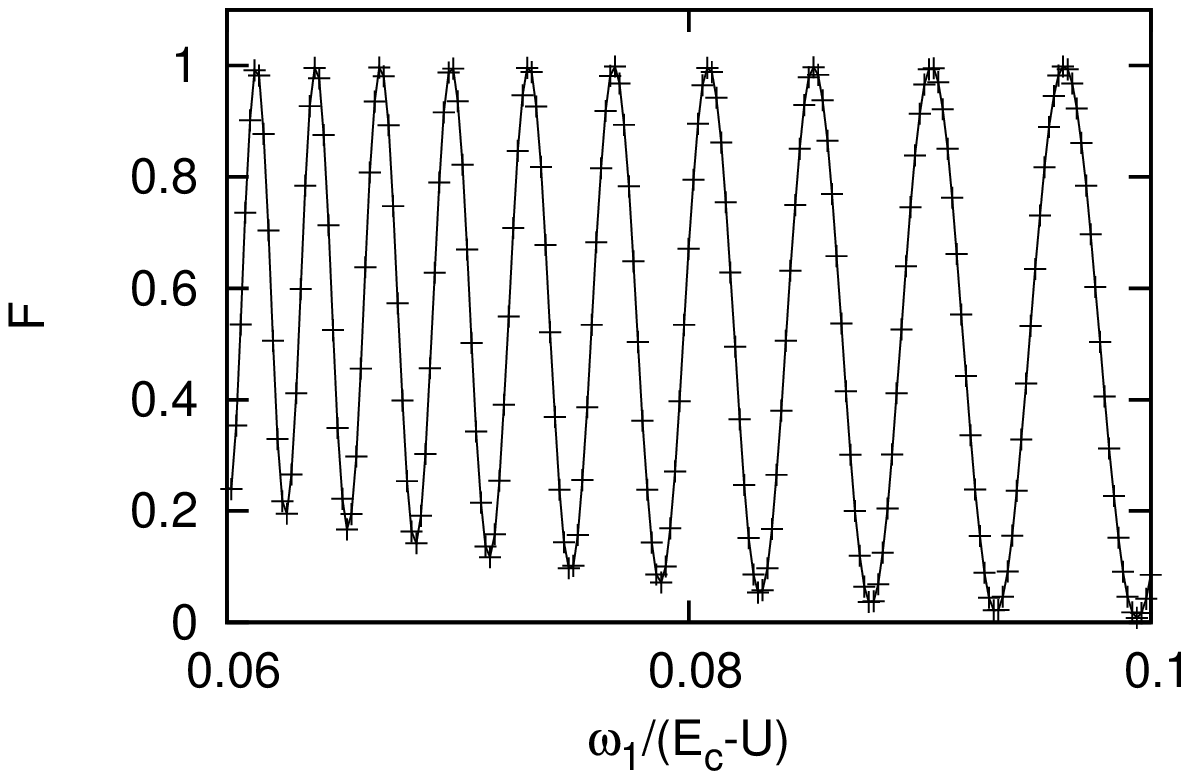}
\includegraphics[width=2.0in,angle=-90]{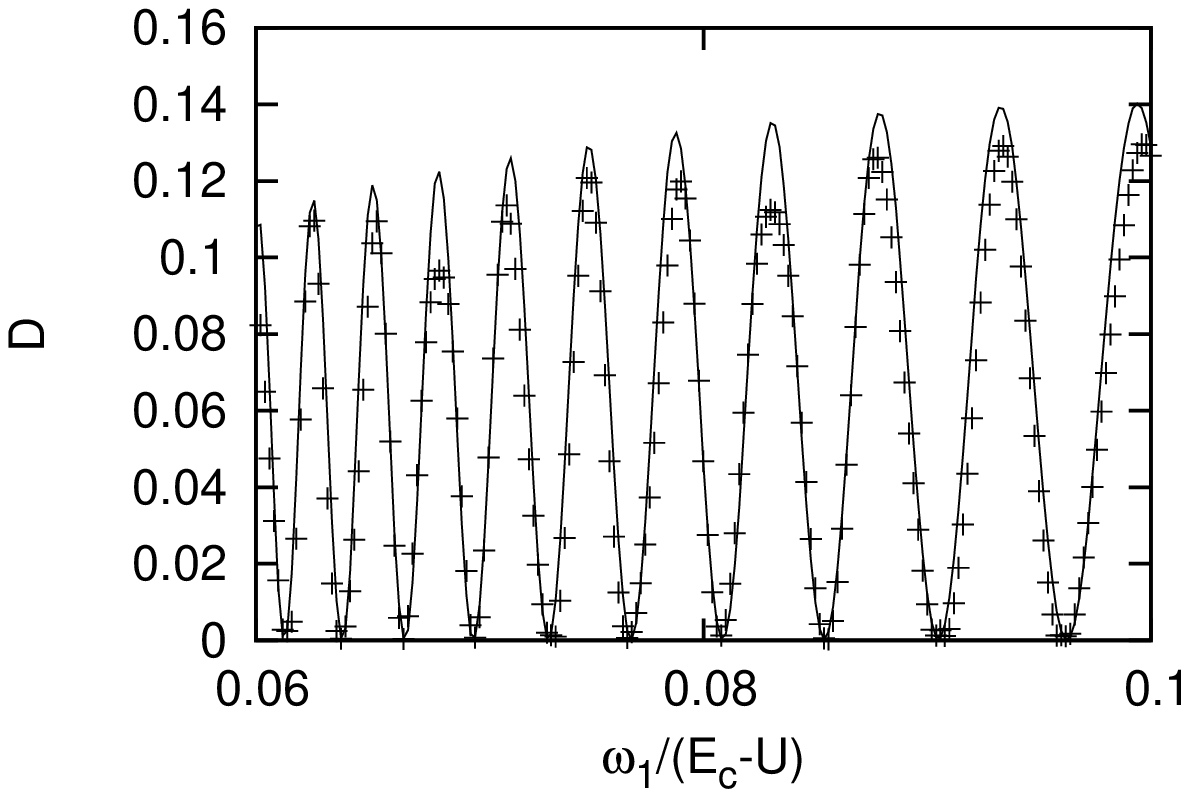}
\caption{Plots of the wavefunction overlap $F$ (top) and the dipole
excitation density $D$ (bottom) as a function of $\omega_1/(E_c-U)$
showing near perfect match between analytical (Eq.\ \ref{eq_anal})
and exact numerical calculations. The points correspond to numerical
results and the lines to Eq. \ref{eq_anal}.} \label{fig_anal_num}
\end{figure}

The non-monotonic dependence of $D$ and $(1-F)$ as a function of the
drive frequency $\omega_1$ is a reminiscent of the analogous
behavior of the probability of excitation for two level systems
subjected to periodic drives. This behavior originates from the
interference effect between probability amplitude of the two-level
system wavefunction at the ground and the excited state on second
passage through the avoided level crossing during the periodic
drive. Such an interference phenomenon is known as St\"uckelberg
interference \cite{nori1}. However, the system of dipoles at hand is
a many-body system with several energy levels whose number increases
with system size. Thus, it is a priori unclear whether the
oscillatory behavior of $D$ and $F$ observed here can also be
explained in terms of such interference phenomenon between a few
states. To understand this further, we therefore analyze the
wavefunction overlap $|c_{\alpha}^2(t)| = |\langle
\psi(t)|\alpha\rangle|^2$, where $|\psi(t)\rangle$ is the state of
the system after time $t$ and $|\alpha\rangle$ are the eigenstates
of $H(t=0)$. A plot of $|c_{\alpha}(t)|^2$ for some selected
$|\alpha\rangle$ satisfying $|c_{\alpha}(t)|^2 \ge 10^{-5}$ at any
time $t$ during the evolution is shown in Fig.\ \ref{fig4} for two
representative frequencies $\omega_1/|U-E_c|=0.051$ and $0.074$ and
for system size $N=14$. The first of these frequencies correspond to
a dip in $D$ and $(1-F)$ while the second to their peak. We find
that in both cases the system starts in the state $|\alpha\rangle=
|0\rangle$ so that $|c_0(t)|^2 \simeq 1$ until the first passage
through the critical point at $t=t_1$. During the first passage, a
few other $|c_{\alpha}(t)|^2$ develops non-zero value as shown in
Fig.\ \ref{fig4}. This is then followed by a regime $t_1\le t \le
t_2$, where the system is close to the AFM ground state having
maximum probability $|c_{\alpha}|^2$ for $\alpha=842$
which corresponds to the AFM ground state. As the system approaches
$t=t_2$ where it completes its second passage through the critical
point, we find that there is again a transfer of weight between
several states. These features are common for both frequencies.
However, the crucial difference between the two cases lies in the
fact that for $t>t_2$ $|\psi (T)\rangle$ has near perfect overlap
with $|\alpha=0\rangle$ (dipole vacuum ground state) when
$\omega_1/|U-E_c|=0.051$ with $\langle n_d \rangle=0$; in contrast,
for $\omega_1/|U-E_c|=0.074$, it has a substantial overlap with
$|\alpha=15\rangle$ which corresponds to $\langle n_d \rangle \simeq
2.1$. The final state also has a non-zero overlap with
$\ket{\alpha=14}$ having $\langle n_d \rangle=1.1$. This difference
originates from the quantum interference between the several states
which gains a finite probability amplitude during the second passage
of the system through the critical point. We have checked that in
between these two frequencies which corresponds to a maxima or a
minima of $D$ and $1-F$, $|\psi(T)\rangle$ always remain a
superposition of three states. In other words, it is possible to
write
\begin{eqnarray}
|\psi(T)\rangle &\simeq&  c_0(\omega_1) |\alpha=0\rangle +
c_{14}(\omega_1) |\alpha=14\rangle \nonumber\\
&& + c_{15}(\omega_1) \ket{\alpha=15} \label{appexp}
\end{eqnarray}
so that
\begin{eqnarray}
F(T) &\simeq& |c_0(\omega_1)|^2, \nonumber\\
D(T) &\simeq&
(n_{14}|c_{14}(\omega_{1})|^2+n_{15}|c_{15}(\omega_1)|^2)/N,
\label{eq_anal}
\end{eqnarray}
where in the last line we have assumed that the ground state at
$t=T$ is a zero dipole state, and $n_{14}$ and $n_{15}$ are the
number of dipoles in the $\ket{\alpha=14}$ and $\ket{\alpha=15}$,
respectively. We compare $D$ and $F$ obtained from Eq.\
\ref{eq_anal} with the numerical calculations performed in Fig.
\ref{fig_anal_num} and observe a very good agreement between the
two.

\begin{figure}[h]
\includegraphics[width=2.5in,angle=-90]{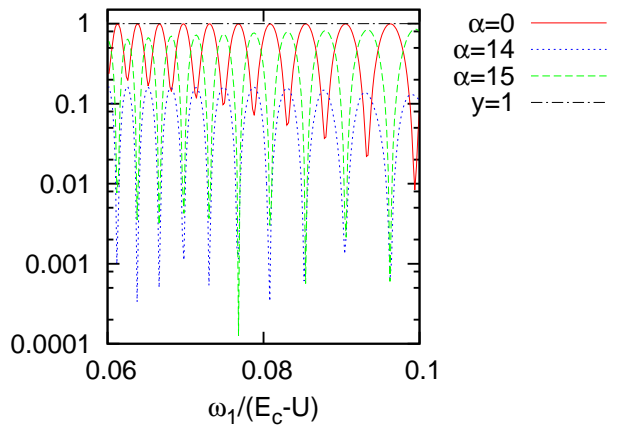}
\caption{Plot of the variation of $|c_{\alpha}|^2$ at $t=T$ as a
function of $\omega_1$ for $\alpha=0,~14$ and $15$. } \label{fig5}
\end{figure}

Thus, we find that the dynamics of the many-body system at the end
of a drive cycle can be described by an effective three-level model
since the wavefunction after the drive is controlled by the
coefficients $c_0(\omega_1)$, $c_{14}(\omega_1)$ and
$c_{15}(\omega_1)$. Numerically, as shown in Fig.\ \ref{fig5}, for a
range of $\omega_1$, these coefficients display oscillatory behavior
as a function of $\omega_1$ which results in the oscillatory
behavior of $D$ and $1-F$. Also, we note that the phenomenon of
dynamic freezing occurs for $\omega_1=\omega^{\ast}$ for which
$|c_0(\omega^{\ast})|^2 \simeq 1$. Hence, we observe a phenomenon
which is a modified version of the St\"uckelberg interference for
the  following reasons. First, similar to the St\"uckelberg
interference phenomenon, the probability of the system to return to
the ground state can be described in terms of a few states (one
needs three states here compared to two states in the usual
descriptions of  St\"uckelberg interference). Second, the occupation
probabilities of these three states display an oscillatory behavior
leading to oscillations of $D$ and $F$. Third, the weight transfer
between these states originates from quantum interference between
several many-body states upon second passage through a quantum
critical point. However, in contrast to the usual two-level systems
where such a phenomenon is first predicted \cite{nori1}, the weight
transfer between the dipole states at the critical points involve
several many-body states; thus the dependence of $c_0$, $c_{14}$ and
$c_{15}$ on $\omega_1$ is determined by interference between
multiple many-body states which, in contrast to the original
St\"uckelberg problem, does not easily admit an analytical
description. However, our analysis at least establishes the fact
that the near-perfect dynamic freezing observed in this system
originates from quantum interference between many-body states during
multiple (two) passages of the system through the PM-AFM quantum
critical point. We note in passing that we have checked that
qualitatively similar phenomenon occurs for other values of
$\omega_1$ corresponding to maxima or minima of $D$ and $1-F$ and
for other system sizes $N=12,16, \, {\rm and}\, 18$.

\subsection{Two-rate Protocol}
\label{trcol}

In this section, we study the dynamics of the system in the presence
of the two-rate protocol given by Eq.\ \ref{prototwo}. We note that
recently such a protocol has been shown to provide a mechanism for
defect suppression in Ref. \onlinecite{sau1} for integrable models.
In the case of integrable models, the two parameters of the
Hamiltonian were varied with rates $\omega_1$ and $\omega_2$ so that
the system crosses the quantum critical point at some time $t_c$. It
was shown that for these models, one of the time dependent
parameters controlled the proximity of the system to the quantum
critical point whereas the other controlled the dispersion of the
quasiparticle at the critical point with rate $\omega_2$. It was also
shown that the defect density $D$ and the residual energy $Q$ for a
$d-$dimensional system obey novel universal power-law behavior given
by \bea D &\sim& \omega_1^{\frac{(2z\nu+1)d}{z(1+z\nu)}}
\omega_2^{-d/z},\,\,  Q \sim
\omega_1^{\frac{(2z\nu+1)(d+z)}{z(1+z\nu)}} \omega_2^{-(d+z)/z}
\label{eq_sau} \eea where $\nu$ and $z$ are the critical exponents
related to correlation length and correlation time, respectively.
Note that both $D$ and $Q$ decrease with $\omega_2$.

We now apply the two-rate protocol to the tilted Bose Hubbard model.
The drive protocol is given by Eq.\ \ref{prototwo}. We start from an
initial PM ground state corresponding to $(U-E_i)/\Delta J=100$ at
$t=0$, with $\Delta E/\Delta J \simeq 200$, and $\epsilon/\Delta J =
0.001$. Both $E(t)$ and $J(t)$ are varied with two different
velocities till the critical point at time $t_c$ which is given by
Eq.\ \ref{tcdef}. The crucial difference of the present case which
constitutes an example of a non-integrable model is that the
microscopic parameters $J$ and $E$ differ from those of the
effective theory controlling the low-energy dynamics.

To identify the parameters of the effective theory which controls
the proximity to the critical point and the velocity or dispersion
of the quasiparticle in this model, we define the instantaneous
quasiparticle gap $\Delta(t)=U-E(t)+\mu_0 J(t)$ in the PM phase. We
note that $\Delta(t_c)=0$. Expanding $\Delta(t)$ around $t=t_c$, one
finds $|\Delta(t)| \sim |\omega_0 (t-t_c)|$, where $\omega_0$ is
given by Eq.\ \ref{tcdef}. Thus, we find that it is $\omega_0$ (and
not $\omega_1$ or $\omega_2$) which controls the proximity to the
critical point. Next, we identify the term which controls the
dispersion. As shown in Ref. \onlinecite{sachdev1}, the
instantaneous velocity of the dipoles in the PM phase is given by
$J(t)^2/|U-E(t)|$. Thus, the velocity of quasiparticles around the
critical point can be estimated to be $v \sim J(t_c) \propto
(\omega_2/\omega_0)$. We have verified our estimates for the gap and
the velocity numerically by studying the gap at the critical point
for a finite system. Identifying $J(t_c)$ as the quasiparticle
velocity near the quantum critical point, and following the
arguments in Ref. \onlinecite{sau1}, it is then straightforward to
obtain \bea D \sim \omega_0^{3/2} \omega_2^{-1}~{\rm and}~ Q \sim
\omega_0^3 \omega_2^{-2}. \eea In terms of the experimental
frequencies $\omega_1$ and $\omega_2$, one thus expects
\begin{eqnarray}
D &\simeq& (\omega_1-\nu_0 \omega_2)^{3/2} \omega_2^{-1} \nonumber\\
Q &\simeq& (\omega_1-\nu_0 \omega_2)^{3} \omega_2^{-2}
\label{micro}
\end{eqnarray}
where $\nu_0/\Delta J = \mu_0 /\Delta E$. For $\nu_0 \ll
\omega_1/\omega_2$, $\omega_0 \simeq \omega_1$ and one recovers the
scaling relations of Eq.\ \ref{eq_sau}.

\begin{figure}[t]
\includegraphics[height=3.4in,angle=-90]{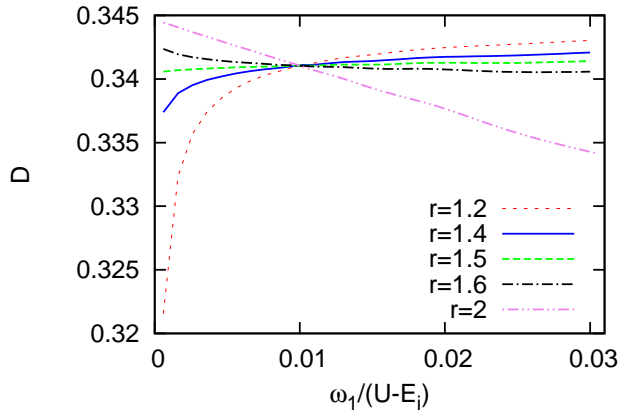}
\caption{Plot of the variation of the dipole excitation density $D$
with $\omega_1$ for different $r$ with fixed $\omega_2=\omega_1^r$.
The plot shows a clear crossover in behavior of $D$ from increasing
to decreasing function of $\omega_1$ as $r$ passes through $3/2$.
Here we have set $\nu_0=0.01$.} \label{fig_diffr}
\end{figure}

We now present numerical results obtained by solving Eq.\
\ref{cprimeeq}. To check the predictions outlined in Eq.\
\ref{micro}, we first set $\omega_2=\omega_1^r$. In this case,
$D\sim \omega_1^{3/2}(1-\nu_0 \omega_1^{r-1})^{3/2} \omega_1^{-r}$.
Once again, for $\nu_0 \ll \omega_1^{1-r}$,
Eq.\ \ref{micro} predicts a crossover in variation of $D$ as a
function of $\omega_1$.
For $r<3/2$, $D$ should increase with $\omega_1$ whereas
it is expected to decrease with $\omega_1$ for $r>3/2$.
On the other hand, at $r=3/2$, $D$ should be a universal number which is
independent of $\omega_1$. A plot of $D$ vs $\omega_1$, shown in
Fig.\ \ref{fig_diffr} for $n_0=1$, confirms this behavior for
different $r$. We note that the decrease of $D$ with $\omega_1$ for
$r>3/2$ shows that it is possible to realize a near-adiabatic
protocol by tuning microscopic parameters $J$ and $E$ in a
non-integrable quantum many-body system.

\begin{figure}[t]
\includegraphics[height=2.2in]{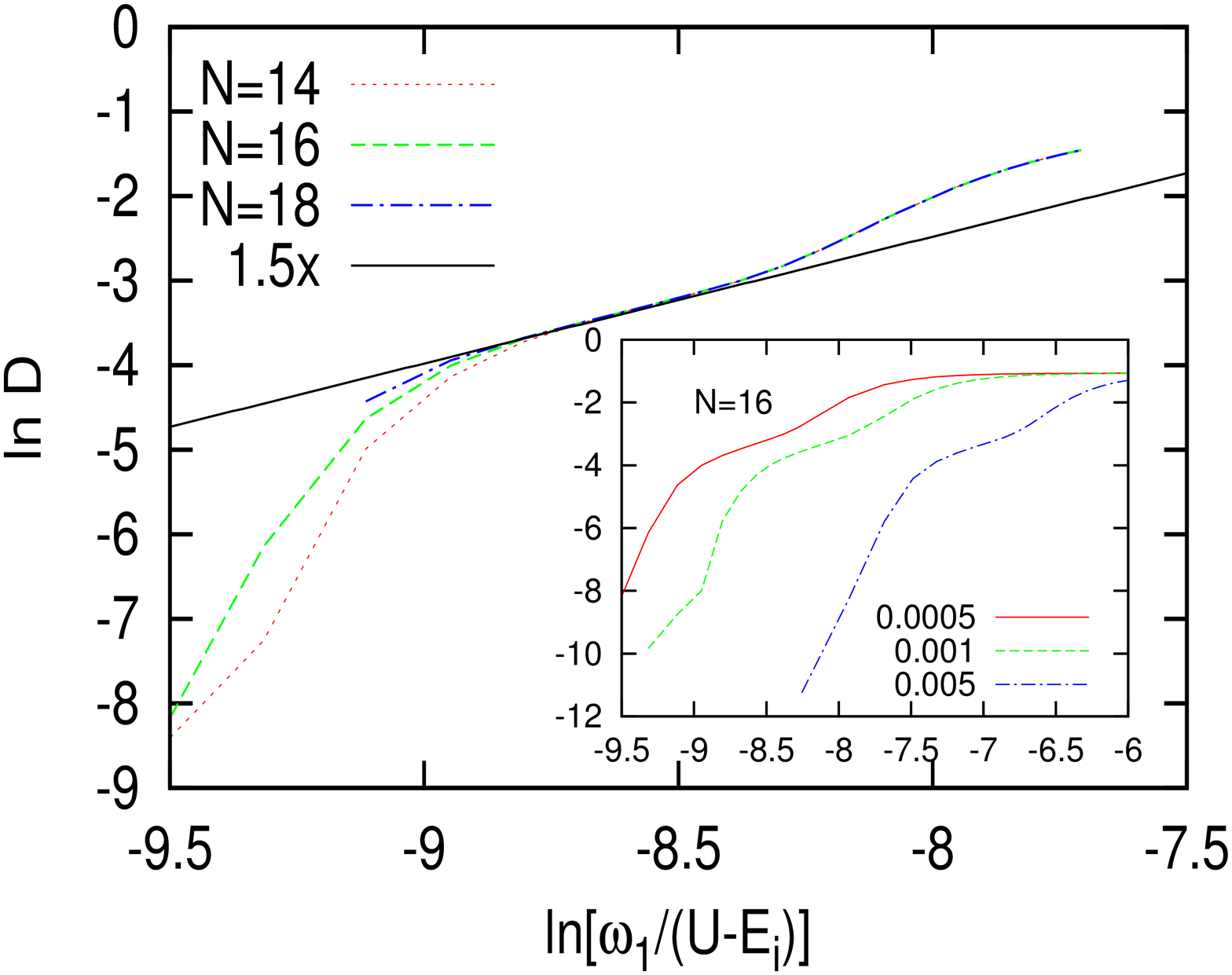}
\caption{Plot of $\ln(D)$ with $\omega_1$ for different system sizes
and fixed $\omega_2/(U-E_i) = 0.0005$. The ramp of $E$ and $J$
starts from $U-E_i=100$ to $U-E_f$ where $E_f$ is the value of the
effective electric field at the critical point as discussed in the
text. The scaling behavior of $D$ predicted in Eq.\ \ref{micro} is over
a finite frequency range $-8.4 \ge \ln(\omega_1/(U-E_i)) \ge -9.0$;
see text for details. Inset: Plot of $\ln(D)$ as a function of
$\omega_1$ for $\omega_2/(U-E_i) = 0.0005$ (red solid line), $0.001$
(green dashed line) and $0.005$ (blue dash-dotted line) with system
size set to $N = 16$. All other parameters are same as in Fig.\
\ref{fig_diffr}.} \label{fig_D_w1}
\end{figure}

\begin{figure}[t]
\includegraphics[height=3.2in,angle=-90]{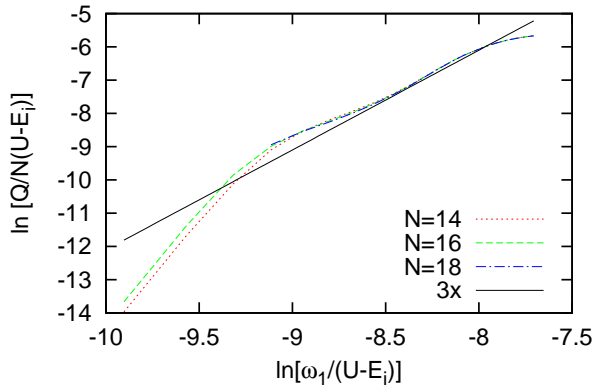}
\caption{Plot of $\ln[Q/(N(U-E_i))]$ with $\omega_1$ for different
system sizes and fixed $\omega_2/(U-E_i)=0.0005$ demonstrating the
agreement between the theoretically predicted $Q \sim \omega_1^{3}$
and the numerical simulations over a range of $\omega_1$ (see text
for details). All other parameters are same as in Fig.\
\ref{fig_diffr}.} \label{fig_Q_w1}
\end{figure}

Next, we study the dynamics of the system keeping $\omega_1$ and
$\omega_2$ independent so that the predicted behavior in Eq.\
\ref{micro} with each of them can be verified. First, we fix
$\omega_2$ and plot the variation of $\ln D$ and $\ln(Q/N)$ as a
function of $\ln[\omega_1/(U-E_i)]$ for different system sizes $18
\ge N\ge 14$ and for several representative values of $\omega_2$ in
Figs.\ \ref{fig_D_w1} and \ref{fig_Q_w1}. These plots show the
expected increase of $D$ and $Q$ as a function of $\omega_1$. We
note from these plots that both $D$ and $Q$ follow the expected
Kibble-Zurek scaling behavior for a finite intermediate range of
$\omega_1$ whose value depend on $\omega_2$. For example, for
$\omega_2=0.05$ in Fig. \ref{fig_D_w1}, we find this range to be
$-8.4 \ge \ln(\omega_1/(U-E_i)) \ge -9.0$. However for lower
$\omega_1$, both $D$ and $Q$ deviates from this scaling behavior.
This can be understood as a finite-size effect. As shown in Ref.\
\onlinecite{kolo1}, the scaling relation for $D$ and $Q$ are
modified by appropriate scaling functions due to finite-sized effect
which leads to Landau-Zener type behavior ($D \sim
\omega_1^{-2}$)for $\omega_1 L \ll 1$. Physically, this can be
understood as the presence of gap in the energy spectrum at the
critical point due to finite-size effect. This gap, which originates
from the presence of a lower momentum cutoff $\sim 1/L$, leads to an
avoided level crossing which leads to Landau-Zener type dynamics for
$\omega L^2 \ll 1$ \cite{kolo1}. For large $\omega_1$, $D$ reaches a
plateau as a function of $\omega_1$ signalling the setting in of the
sudden quench regime where the response of the system becomes
independent of $\omega_1$.

\begin{figure}[t]
\includegraphics[height=2.2in]{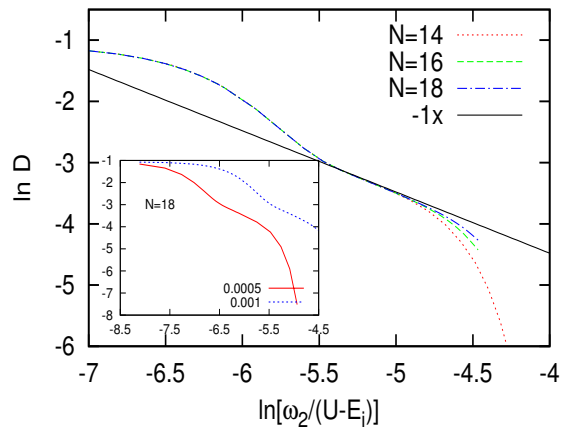}
\caption{Plot of $\ln(D)$ as a function of rate $\omega_2$ for
different system sizes and fixed $\omega_1/(U-E_i)=0.001$ showing
decrease of $D$ with $\omega_2$. The theoretically predicted slope
obtained from the scaling theory is also shown for comparison. The
figure also demonstrates the role of finite size effects; the $N=18$
data follows the power law till larger value of $\omega_2$. Inset:
Plot of $\ln(D)$ as a function of $\omega_2$ for fixed $N=18$ and
$\omega_1/(U-E_i)=0.0005$ (red solid line) and $0.001$ (blue dotted
line). All other parameters are same as in Fig.\ \ref{fig_diffr}.}
\label{fig_D_w2}
\end{figure}

Finally, we present the numerical results for the dependence of $D$
on $\omega_2$ as shown in Fig. \ref{fig_D_w2}. We again find agreement
between the theoretically expected behavior $D \sim \omega_2^{-1}$ (note
that we have set $\nu_0 \ll 1$) for a range of drive frequency $-5.5
\le \ln(\omega_2/(U-E_i)) \le -4.7$. As also found for $\omega_1$
dependence of $D$, scaling behavior does not hold for smaller
frequencies $\ln(\omega_2/(U-E_i)) \le -5.5$ suggesting setting in
of finite size effects. For larger frequencies
$\ln(\omega_2/(U-E_i)) \ge -4.7$, $D$ registers a sharper drop than
$\omega_2^{-1}$ suggesting the end of scaling regime. We note that
for the entire range, $D$ is a monotonically decreasing function of
$\omega_2$ which indicates larger excitation suppression with
increasing $\omega_2$. We have checked that $F$ and $Q$ show
qualitatively similar behavior as a function of $\omega_2$.

\section{Discussion}
\label{conc}

In this work, we have studied the behavior of bosons in a tilted
one-dimensional optical lattice in the presence of both periodic and
two-rate drives using exact diagonalization and for finite size
systems $N\le 18$. For the periodic drive protocol which takes the
system twice through the intermediate critical point separating the
paramagnetic (dipole vacuum) and the ferromagnetic (maximal dipole)
states, we have demonstrated the presence of non-monotonic
dependence of the dipole excitation density $D$ and the defect
density $1-F$ (where $F$ is the wavefunction overlap between the
final state after the drive and the initial ground state) as
measured at the end of a complete drive cycle. We have shown that
such a behavior originates from quantum interference between the
wavefunctions of different states of the boson Hilbert space and
constitutes a many-body generalization of the St\"uckelberg
interference phenomenon for two-level systems. Our work also
identifies special frequencies where such an interference phenomenon
leads to near zero values of $D$ and $1-F$; at these frequencies the
system exhibits a near-perfect dynamic freezing $\ln(1-F) \sim -8$
in the sense that the system wavefunction, at the end of a drive
period, has a near perfect overlap with the starting ground state
wavefunction. We note that such an interference phenomenon has no
analog in ramp \cite{kolo1} or quench \cite{sengupta1} dynamics of
the models studied earlier.

For the two-rate protocol, which constitutes a ramp of both the
electric field $E$ and the hopping amplitude $J$ of the bosons
taking the system from the paramagnetic (dipole vacuum) phase to the
critical point, we demonstrate suppression of dipole excitation
density $D$ as a function of $\omega_2$. We demonstrate that $D$ is
a monotonically decreasing function of $\omega_2$ and chart out the
scaling regime where $D \sim \omega_2^{-1}$ for a fixed $\omega_1$.
We also study the behavior of the system by setting $\omega_2
=\omega_1^r$ and demonstrate that the system exhibits a crossover at
$r=3/2$. For $r< 3/2$, $D$ increases with $\omega_1$ while it
decreases for $r>3/2$; at $r=3/2$ $D$ is a constant. We note that our results
constitutes an experimentally realizable demonstration for defect
suppression on passage of a many-body system through a quantum
critical point.

Finally, we discuss possible experiments which can test our theory.
In this respect, we note that tilted experimental lattice systems
has already been experimentally studied for $N=12$ in Ref. \onlinecite{bakr1}; in
particular, the ground state phase diagram of the model has been
experimentally verified using direct measurement of on-site parity
of occupation of the bosons. More recently, other techniques which
allows for direct measurement of boson occupation at a given site
has also been reported \cite{mukund1}. Our suggested experiments are
build on these and are as follows. First, we suggest measurement of
dipole density for the periodically driven titled lattice system
where the effective electric field is varied periodically as a
function of time. Such an electric field is realized in experiments
by using a spatially varying Zeeman field; consequently, its
periodic variation can be achieved by making the Zeeman field a
periodic function of time. We suggest periodic tuning of the electric
field from a value $E_i$ which corresponds to the dipole vacuum
state (or the $n=1$ Mott phase of the bosons)
through the phase transition value $E_c$ followed by subsequent
measurement of number of sites, $n_{\rm even}$, with even boson
occupation number (which corresponds to the dipole density $D$) at
the end of a period of the drive. Our theoretical prediction is that
$n_{\rm even}$ shall be a periodic function of the drive frequency
$\omega_0$. We also predict the existence of specific values of
$\omega_0$ where the system shall exhibit near-perfect dynamic
freezing leading to $n_{\rm even} \to 0$. Second, we suggest a
linear ramp protocol for variation of the hopping amplitude of these
bosons $J$ (with rate $\omega_2$) and the effective electric field
$E$ (with rate $\omega_1$) which takes the system from the dipole
vacuum phase to the critical point. We note that such a protocol can
be achieved by simultaneous linear ramp of the applied Zeeman field
and the laser intensity controlling the depth of the optical
lattice. For this protocol, we predict $n_{\rm even}$ at the end of
the ramp will be a monotonically decreasing function of $\omega_2$
for a fixed $\omega_1$.

In conclusion, we have studied the response of bosons in a tilted
optical lattice in the presence of periodic and two-rate protocols.
For the periodic protocol, we have identified special frequencies at
which the system exhibits dynamic freezing and have related this
phenomenon to a many-body version of St\"uckelberg interferece
effect. For the two-rate protocol, we have identified drive
frequency ranges at which the finite-size systems displays scaling
behavior as predicted by Kibble-Zureck theory. We have also
demonstrated that an increase of $J$ with a rate $\omega_2$ leads to
decrease of $D$ and $Q$ leading to realization of a near-adiabatic
protocol for this system on passage through a quantum critical
point. We have also suggested concrete experiments which can test
our theory.

\begin{acknowledgments}
UD gratefully acknowledges funding from DST-INSPIRE Faculty
fellowship by DST, Govt. of India, and the hospitality of IACS, Kolkata,
during her visits.
\end{acknowledgments}

\end{document}